\documentclass[pre,
   superscriptaddress,
   twocolumn,
   eqsecnum,
   showpacs,
   preprintnumbers,
   byrevtex]{revtex4-1}
\usepackage{amsfonts}
\usepackage{amssymb}
\usepackage{amsmath}
\usepackage{color}
\usepackage{graphicx}
\usepackage{subfigure}
\usepackage[pdftex,colorlinks=true]{hyperref}
\newcommand{\um}{u_m}
\newcommand{\uI}{u^{(i)}_m}
\newcommand{\uII}{u^{(ii)}_m}
\newcommand{\uIII}{u^{(iii)}_m}
\newcommand{\uIV}{u^{(iv)}_m}
\newcommand{\uV}{u^{(v)}_m}
\newcommand{\uVI}{u^{(vi)}_m}
\newcommand{\uVII}{u^{(vii)}_m}
\newcommand{\uVIII}{u^{(viii)}_m}
\newcommand{\uIX}{u^{(ix)}_m}

\newcommand{\del}{\Delta x}

\begin{document}
%
%
%

\preprint{LA-UR 10-05872}

\title
   {Stability and dynamical properties of Cooper-Shepard-Sodano compactons}

\author{Bogdan Mihaila}
\email{bmihaila@lanl.gov}
\affiliation{Materials Science and Technology Division, 
Los Alamos National Laboratory, Los Alamos, New Mexico 87545, USA} 

\author{Andres Cardenas}
\email{andres.cardenas@nyu.edu}
\affiliation{Materials Science and Technology Division, 
Los Alamos National Laboratory, Los Alamos, New Mexico 87545, USA}   
\affiliation{Physics Department, New York University, New York, NY 10003, USA}

\author{Fred Cooper}
\email{cooper@santafe.edu}
\affiliation{Santa Fe Institute, Santa Fe, NM 87501, USA}
\affiliation{Theoretical Division and Center for Nonlinear Studies, 
Los Alamos National Laboratory, Los Alamos, New Mexico 87545, USA}

\author{Avadh Saxena}
\email{avadh@lanl.gov}
\affiliation{Theoretical Division and Center for Nonlinear Studies, 
Los Alamos National Laboratory, Los Alamos, New Mexico 87545, USA}
   
\begin{abstract}
Extending  a Pad\'e  approximant method used for studying compactons in the Rosenau-Hyman (RH) equation, we study the numerical  stability  of single compactons of the Cooper-Shepard-Sodano (CSS) equation  and their pairwise interactions.  The CSS equation has  a conserved Hamiltonian which has allowed various approaches for studying analytically the nonlinear stability of the solutions.  We study three different compacton solutions and find they are numerically stable. Similar to the collisions between RH compactons, the CSS compactons reemerge with same coherent shape when scattered. 
\textcolor{black}{
The time evolution of the small-amplitude ripple resulting after scattering depends on the values of the parameters $l$ and $p$ characterizing the corresponding CSS equation.
}
The simulation of the CSS compacton scattering requires a much smaller artificial viscosity to obtain numerical stability than in the case of RH compacton propagation. 
\end{abstract}


\maketitle

\section{Introduction}

Following their discovery \cite{RH93}, compactons, or solitary waves defined on a compact support, have found diverse applications in physics~\cite{LD98,BP96}, ocean dynamics~\cite{GOSS98}, magma dynamics~\cite{SSW07,SWR08}, mathematical physics~\cite{brane,KG98,CDM98}, nonlinear lattice dynamics~\cite{SMR98,C02,CM06,PKJK06,PKJK06,RP05,PR06,R00}, and medicine~\cite{KEOK06}. Multidimensional compactons have also been discussed in~\cite{R06,RHS07}, and compact structures have been studied \textcolor{black}{in the context of the discrete Burridge-Knopoff model~\cite{ComteBK}, and in the context of discrete and continuous  Klein-Gordon models~\cite{ComteKG,RK08,RK10}}. A recent review of nonlinear evolution equations with cosine/sine compacton solutions can be found in Ref.~\onlinecite{RV09}.

The $K(l,p)$ compactons discussed first by Rosenau and Hyman (RH) are examples of a class of traveling-wave solutions with compact support  resulting from the balance of both nonlinearity and nonlinear dispersion. RH discovered these compactons in their  studies of pattern formation in liquid drops using a family of fully nonlinear Korteweg-de Vries (KdV) equations~\cite{RH93},
\begin{equation}
   u_t + (u^l)_x + (u^p)_{xxx} = 0
   \>,
   \label{eq:kmn}
\end{equation}
where $u \equiv u(x,t)$ is the wave amplitude, $x$ is the spatial coordinate and $t$ is time. Equation~\eqref{eq:kmn} is known as the $K(l,p)$ compacton equation. The RH compactons have the remarkable soliton property that after colliding with other compactons they reemerge with the same coherent shape. The collision site is marked by the creation of a compact ripple. \textcolor{black}{The positive- and negative-amplitude parts of the ripple decay slowly into low-amplitude compactons and anti-compactons, respectively~\cite{RH93}.}
De Frutos \emph{et al.} showed~\cite{dF95}, and Rus and Villatoro confirmed recently~\cite{RV07a}, that shocks are generated during compacton collisions. 

In general, Eq.~\eqref{eq:kmn} does not exhibit the usual energy conservation law. Therefore, Cooper, Shepard and Sodano (CSS) proposed a different generalization of the KdV equation based on the first-order Lagrangian~\cite{CSS93}
\begin{equation}
   L(l,p) = \int \Bigl [ \frac{1}{2} \phi_x \phi_t + \frac{(\phi_x)^l}{l(l-1)}
   - \alpha (\phi_x)^p (\phi_{xx})^2 \Bigr ] \, dx
   \>,
   \label{eq:Llp}
\end{equation}
which leads to the equation:
\begin{equation}
u_t + u^{l-2} u_x -  p [u^{p-1} (u_x)^2]_x + 2 \alpha [ u^p u_x]_{xx} =0 \>. \label{eq:css}
\end{equation} 
Here, we have $u=\phi_x$.
Since then, various other Lagrangian generalizations of the KdV equation have been considered~\cite{AC93,CHK01,DK98,CKS06,BCKMS09}. 
The equation for the solitary waves  is obtained by substituting $u(x,t) = f(x-ct) \equiv f(y)$ into Eq. (\ref{eq:css}) and then integrating twice and setting the integration constants to zero.  One obtains:
\begin{equation}
\frac{c}{2} f^2 - \frac{f^l}{l(l-1)} + \alpha (f')^2 f^p =0 \>.
\label{eq:css_c}
\end{equation}
\textcolor{black}{
Anti-compacton solutions correspond to the transformation $f \rightarrow -f$. Therefore, from Eq.~\eqref{eq:css_c}, we find that for anti-compactons to exist, $l-p$ must be an even integer. Moreover, when $p$ is odd $c$ changes sign and the anti-compacton travels with negative velocity, whereas for $p$ even the velocity of the anti-compacton is positive.}

Compacton solutions are constructed by patching a compact portion of a periodic solution that is zero at both ends to a solution that vanishes outside the compact region to  give a weak solution to the equation.  We see that for there to be a solution of that type, 
$p \leq 2$ and $l \geq p$.  The condition for a weak solution is that the jump across the boundary of the equation of motion at $x_0$ where $f[x_0]$=0 is zero. That is 
\begin{equation}
Disc[ (f')^2 f^p ]_{x_0} = 0 \>.
\end{equation}
This is always satisfied if there is no infinite jump in the derivative of the function.  The stability analysis of 
the solutions relies on the fact that the equation of motion for $f(y)$  can be obtained from an Action functional: 
\begin{equation}
\Phi[f] =  \int dy \left[ \frac{c}{2} f^2 - \frac{f^l}{l(l-1)} +\alpha (f')^2 f^p \right] \>.
\end{equation}
We recognize this functional as the value of 
\begin{equation}
P[f]c + H[f] \>,
\end{equation}
where $P[f]$ and $H[f]$ are the values of the conserved momentum and Hamiltonian respectively for the solitary wave $f$.
The once integrated equation of motion for the solitary wave is obtained from the equation
\begin{equation}
\frac{\delta \Phi}{\delta f} = 0 \> ,
\end{equation}
and the unintegrated  equation of motion for $f[y]$ is given by 
\begin{equation}
\partial_y  \frac{\delta \Phi}{\delta f} = 0 \>.
\end{equation}


For the RH equation, stability of the compacton has been demonstrated numerically as well as by a linear stability analysis of the radiation induced by the numerical method \cite{RV07b}.  
For the CSS equation, because of the existence of a Hamiltonian formulation, various other methods of studying nonlinear stability have been explored  such as Lyapunov stability \cite{DK98,Lyapunov}  and stability of the solutions under scale transformations \cite{DK98,Derrick}. 
However, apart from a numerical study of the evolution and scattering of the compactons in the generalized CSS equation by  Cooper, Khare and Hyman \cite{CHK01} using pseudospectral methods, there has been no systematic study until now of the stability of the compacton solutions to the  CSS equation.  Nor has there been any study of whether the solutions that arise from a Hamiltonian dynamical system behave differently from those obeying the four conservation laws of the RH equation \cite{RH93}  (without energy conservation).  It is this gap in our knowledge that we hope to fill by this study.

To study stability under scale transformations it is sufficient to study the change in the Hamiltonian for fixed momentum $P$ \cite{DK98}. 
That is we let 
\begin{equation}
f(x)  \rightarrow \beta^{1/2} f(\beta x) ,
\end{equation}
which leaves $P= \int dx f^2/2$ unchanged. 
The Hamiltonian 
\begin{equation}
H= \int dx \Bigl[ \alpha  f^l (f')^p - \frac{1}{l(l-1)}f^l \Bigr]  \equiv H_1 - H_2 \>,
\end{equation}
is then transformed into 
\begin{equation}
H_1(\beta) = \beta^{\frac{1}{2}(l+3p-2)} H_1 - \beta^{\frac{1}{2}(l-2)}  H_2 \>. 
\end{equation}

The exact solution satisfies:
\begin{equation}
\frac{\partial H}{\partial \beta}\Bigr |_{\beta =1} = 0 \>. 
\end{equation}
This yields 
\begin{equation}
 (l-2) H_2 = (p+4) H_1\>.
\end{equation}
 The second derivative at $\beta =1$ can then be written as
 \begin{equation}
\frac{\partial^2 H}{\partial \beta^2 }\Bigr |_{\beta =1} =  \frac{1}{4} (p+4)(p-l+6)  H_1 \>. \nonumber \\
\end{equation}
Since $H_1$ and  $H_2$ are positive definite we find that the solutions are stable  to a small scale transformation when 
\begin{equation}
2 < l < p+6 \>.
\end{equation}
This includes all the solutions we will be studying here. 

In a recent paper~\cite{pade_paper}, we performed a systematic derivation of a Pad\'e approximants method \cite{baker} for calculating  derivatives of smooth functions on a uniform grid  by deriving higher-order approximations using traditional finite-differences formulas. Our derivation contained as special cases the Pad\'e approximants first introduced by Rus and Villatoro~\cite{RV07a,RV07b,RV08}.  
We note that the $ L(l.p)$ compactons feature higher-order nonlinearities and terms with mixed-derivatives that are not present in the $K(p,p)$ equations. Therefore, in this paper we will extend our earlier approach~\cite{pade_paper}, so that we can study the compactons that occur in the CSS equation. This approach can also be applied to the recent $\mathcal{PT}$ generalizations of that equation~\cite{BCKMS09}.


This paper is outlined as follows.  In Sec.~\ref{sec:pade}, we review briefly the main findings with respect to the numerical schemes based on Pad\'e approximants derived in Ref.~\onlinecite{pade_paper}. Our numerical approach to solving the CSS equation is described in Sec.~\ref{sec:numerics}. 
In section~\ref{sec:res} we study numerically the stability of several compacton solutions of the  CSS equation, and we also study the pairwise interactions of these compactons.  We compare our results on stability with our previous numerical study of the $K(2,2)$ equation~\cite{pade_paper}. 
We summarize our main findings in Sec.~\ref{sec:concl}.

%
%

\section{Pad\'e approximants}
\label{sec:pade}

We consider a smooth function $u(x)$, defined on the interval $x \in [0,L]$, and discretized on a uniform grid, $x_m = m \, h$, with $m=0,1,\cdots,M$, and $h=L/M$. 
Pad\'e approximants of order $k$ of the derivatives of $u(x)$ are defined as \emph{rational} approximations of the form
\begin{align}
   \uI & \ = \frac{\mathcal{A}(E)}{\mathcal{F}(E)} \ \um + \mathcal{O}(\del^k)
   \>,
   \\
   \uII & \ = \frac{\mathcal{B}(E)}{\mathcal{F}(E)} \ \um + \mathcal{O}(\del^k)
   \>,
   \\
   \uIII & \ = \frac{\mathcal{C}(E)}{\mathcal{F}(E)} \ \um + \mathcal{O}(\del^k)
   \label{eq:c1}
   \>,
   \\
   \uIV & \ = \frac{\mathcal{D}(E)}{\mathcal{F}(E)} \ \um + \mathcal{O}(\del^k)
   \>,
\end{align}
where we have introduced the \emph{shift} operator, $E$, as
\begin{align}
   E^k \, \um = u_{m+k} \>.
\end{align}
Even- and odd-order derivatives require approximants that are symmetric and antisymmetric in~$E$, respectively. The familiar second-order accurate approximation of derivatives based on finite-differences are trivial examples of Pad\'e approximants
\begin{align}
   \mathcal{A}_1(E)
   & \ = \frac{1}{2\del} \Bigl [ E - E^{-1} \Bigr ]
   \>,
   \\
   \mathcal{B}_1(E)
   & \ = \frac{1}{\del^2} \Bigl [ E - 2 + E^{-1} \Bigr ]
   \>,
   \\
   \mathcal{C}_1(E)
   & \ = \frac{1}{2\del^3} \Bigl [ E^2 - 2 E + 2 E^{-1} - E^{-2} \Bigr ]
   \>,
   \\
   \mathcal{D}_1(E)
   & \ = \frac{1}{\del^4} \Bigl [ E^2 - 4 E + 6 - 4 E^{-1} + E^{-2} \Bigr ]
   \>,
\end{align}
corresponding to $\mathcal{F}_1(E) = 1$~\cite{RV07b}. Still keeping $\mathcal{F}_1(E) = 1$, but incorporating the additional grid points, $\{x_{m\pm2} \}$, we can obtain fourth-order accurate approximation for the derivatives $\uI$ and $\uII$, as
\begin{align}
\label{til_A1}
   \mathcal{\tilde A}_1(E) & \ =
   - \,
   \frac{1}{12 \del} \,
   \Bigl [
      E^2 - 8 E + 8 E^{-1} - E^{-2}
   \Bigr ]
   \>,
   \\ 
   \mathcal{\tilde B}_1(E) & \ =
   - \, \frac{1}{2 \del^2} \,
   \Bigl [
      E^2 - 6 E + 10 - 6  E^{-1} + E^{-2}
   \Bigr ]
   \>.
\label{til_B1}
\end{align}

Previously~\cite{pade_paper}, we showed on general grounds that the Pad\'e-approximant approach allows one to improve the numerical representations of only three of the four  lowest-order derivatives of  $u(x)$,  when involving only the grid points $\{x_m, x_{m\pm1}, x_{m\pm2} \}$. To obtain a fourth-order accurate approximation of the derivatives, we can either begin by improving the third-order derivative, $\uIII$, or the fourth-order derivative, $\uIV$. Because in the compacton-dynamics problem~\cite{RH93,SC81,dF95,IT98,RV07a,RV07b,LSY04}, the fourth-order derivative enters only through the artificial viscosity term needed to handle shocks, in Ref.~\onlinecite{pade_paper} we chose to improve the approximation corresponding to the third-order derivative, $\uIII$, and focused on obtaining fourth- or higher-order accurate Pad\'e approximants of $\uI$, $\uII$, and $\uIII$, on the subset of grid points, $\{x_m, x_{m\pm1}, x_{m\pm2} \}$.

Using Eqs.~\eqref{eq:c1}, we introduced an operator, $\mathcal{F}(E)$, symmetric in~$E$, as
\begin{align}
   &
   \mathcal{F}(E) \, \uIII
   =
   \frac{1}{a} \,
   \Bigl [ \bigl ( E^2 + E^{-2} \bigr )
            + b \bigl ( E + E^{-1} \bigr )
            + c \Bigr ] \, \uIII
   \>,
\end{align}
such that
\begin{align}
   \mathcal{F}(E) \, \uIII
   =
   \mathcal{C}_1 (E) \, \um
   + \mathcal{O}(\del^k)
   \>,
\end{align}
and showed that for
\begin{align}
   a = 4\tau \>, \quad
   b = \tau-4 \>, \quad
   c = 2(\tau+3) \>,
\end{align}
we obtain
\begin{align}
   \uIII = & \
   \frac{\mathcal{C}_1(E)}{\mathcal{F}(E)} \, \um
   - \uVII \Bigl ( \frac{1}{60} - \frac{1}{\tau} \Bigr ) \, \frac{\del^4}{4}
\label{eq:uIII}
   \\ \notag &
   - \uIX \Bigl ( \frac{43}{2520} - \frac{1}{\tau} \Bigr ) \, \frac{\del^6}{24}
   + \mathcal{O}(\del^8)
   \>.
\end{align}
Correspondingly, the Pad\'e approximant of the first-order derivative, $\uI$, is obtained as
\begin{align}
   \uI = & \
   \frac{\mathcal{A}_2(E)}{\mathcal{F}(E)} \, \um
   \textcolor{black}{-} \uV \Bigl ( \frac{1}{30} - \frac{1}{\tau} \Bigr ) \, \frac{\del^4}{4}
\label{eq:uI}
   \\ \notag &
   \textcolor{black}{-} \uVII \Bigl ( \frac{1}{105} - \frac{1}{4\tau} \Bigr ) \, \frac{\del^6}{6}
   + \mathcal{O}(\del^8)
   \>,
\end{align}
with
\begin{align}
   \mathcal{A}_2(E) & \ =
   \frac{1}{24 \del} \,
   \Bigl [
      E^2 + 10 E - 10 E^{-1} - E^{-2}
   \Bigr ]
   \>,
\end{align}
and the Pad\'e approximant of the second-order derivative, $\uII$, is 
\begin{align}
   \uII = & \
   \frac{\mathcal{B}_2(E)}{\mathcal{F}(E)} \, \um
   \textcolor{black}{-} \uVI \Bigl ( \frac{7}{180} - \frac{1}{\tau} \Bigr ) \, \frac{\del^4}{4}
\label{eq:uII}
   \\ \notag &
   \textcolor{black}{-} \uVIII \Bigl ( \frac{29}{840} - \frac{1}{\tau} \Bigr ) \, \frac{\del^6}{24}
   + \mathcal{O}(\del^8)
   \>,
\end{align}
with
\begin{align}
   \mathcal{B}_2(E) & \ =
   \frac{1}{6 \del^2} \,
   \Bigl [
      E^2 + 2 E - 6 + 2  E^{-1} + E^{-2}
   \Bigr ]
   \>,
\end{align}
and the Pad\'e approximant of the fourth-order derivative, $\uIV$, is
\begin{align}
   \uIV & \ = \frac{\mathcal{D}_1(E)}{\mathcal{F}(E)} \ \um
   \textcolor{black}{+} \uVI \frac{\del^2}{12}
   +
   \mathcal{O}(\del^4)
   \>.
\end{align}

In order to numerically study the stability and dynamical properties of compactons, we will consider a suite of different fourth-order accurate approximation schemes, to make sure that results are independent of the peculiarities of a particular approximation scheme. Therefore, just like in Ref.~\onlinecite{pade_paper}, we will consider here several sets of approximants that mix fourth-order accurate approximations for two of the derivatives $\uI$, $\uII$, and $\uIII$, with a sixth-order accurate Pad\'e approximant for the third one, together with an ``optimal'' fourth-order approximation scheme that minimized the extent of the radiation train in our previous study of $K(2,2)$ compactons. 

{\rm{\textbf{(6,4,4)}} scheme:} This approximation scheme is an extension of the scheme introduced by Sanz-Serna \emph{et al.}~\cite{dF95,SC81}  using a fourth-order Petrov-Galerkin finite-element method, and corresponds to choosing $\tau=30$ in Eqs.~\eqref{eq:uIII} and \eqref{eq:uI}. We have
\begin{align}
   \mathcal{F}_{[644]}(E) & \ =
   \frac{1}{120} \,
   \Bigl [
      E^2 + 26 E + 66 + 26 E^{-1} + E^{-2}
   \Bigr ]
   \>.
\end{align}

{\rm{\textbf{(4,6,4)}} scheme:} A sixth-order accurate approximation for $\uII$, can be obtained by requiring $\tau=180/7$. Then, we have
\begin{align}
   \mathcal{F}_{[464]}(E) & \ =
   \frac{1}{720} \,
   \Bigl [
      7 E^2 + 152 E + 402 + 152 E^{-1} + 7 E^{-2}
   \Bigr ]
   \>.
\end{align}

{\rm{\textbf{(4,4,6)}} scheme:} For $\tau=60$, the coefficient of $\del^4$ vanishes in Eq.~\eqref{eq:uIII} and we obtain a sixth-order accurate approximation for $\uIII$, as
\begin{align}
   \mathcal{F}_{[446]}(E) & \ =
   \frac{1}{240} \,
   \Bigl [
      E^2 + 56 E + 126 + 56 E^{-1} + E^{-2}
   \Bigr ]
   \>.
\end{align}
This scheme is an extension of the scheme introduced first by Rus and Villatoro~\cite{RV07b,RV07a}.

{\rm{\textbf{(4,4,4)}} scheme:} Finally, for $\tau=5$, the smallest value of $\tau$ leading to integer positive values of $a$, $b$, and $c$,  we obtain
\begin{align}
   \mathcal{F}_{[444]}(E) & \ =
   \frac{1}{20} \,
   \Bigl [
      E^2 + E + 16 + E^{-1} + E^{-2}
   \Bigr ]
   \>.
\end{align}
While only leading to a fourth-order accurate approximation scheme, the above choice of $\tau$ was shown to minimize the extent of the radiation train in our previous study of $K(2,2)$ compactons.

%
%

\begin{figure*}[t]
   \subfigure[K(p,p) $u_c$: $p=2$.]
      { \includegraphics[width=\columnwidth]{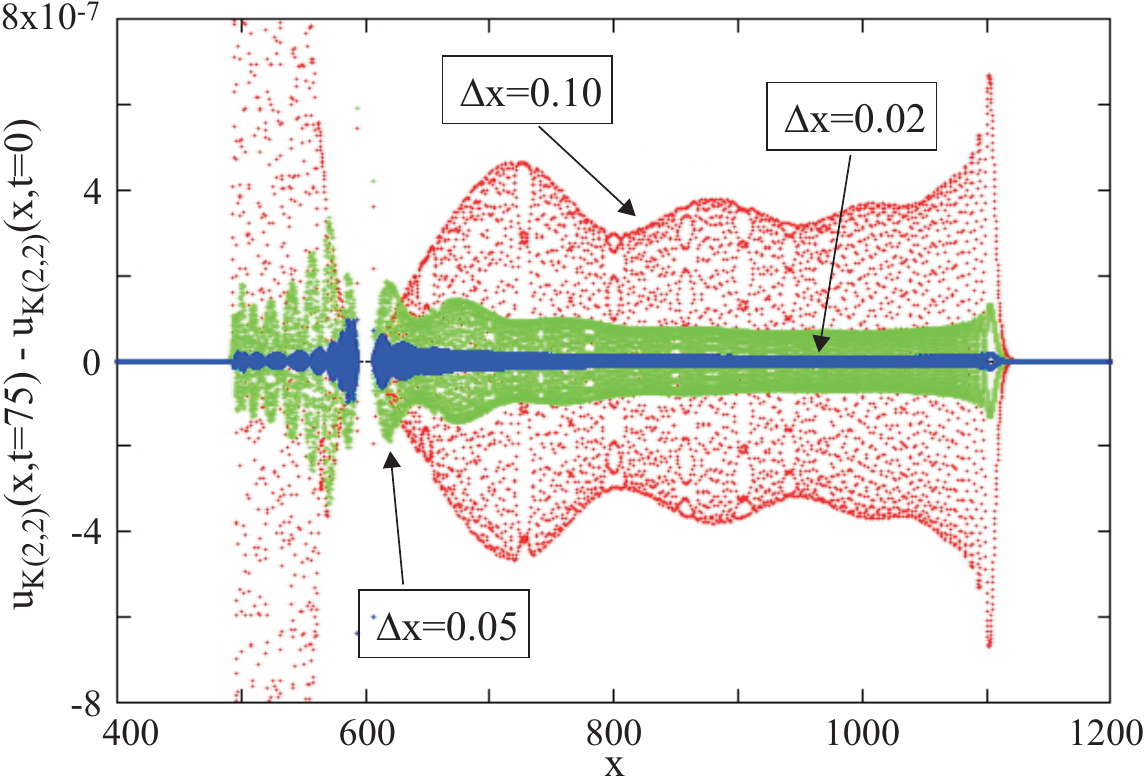} }
   \subfigure[CSS~$u_1$: $p=1,l=3$]
      { \includegraphics[width=\columnwidth]{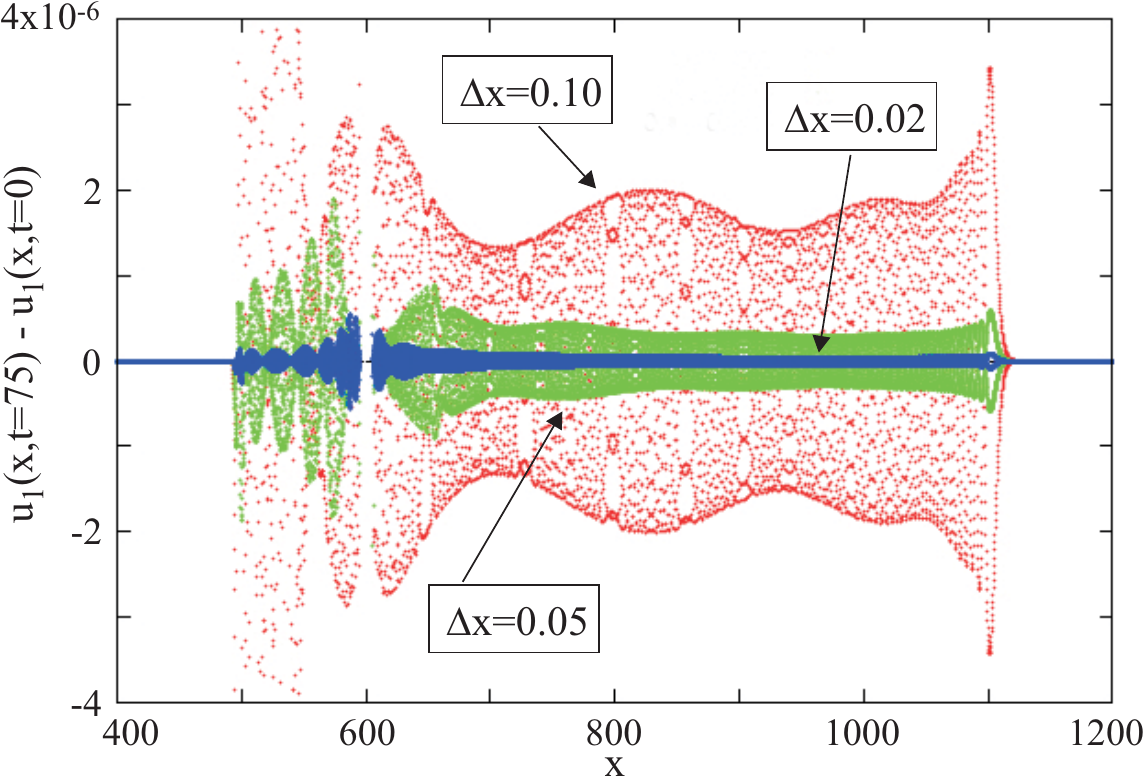} }
   \subfigure[CSS~$u_2$: $p=2,l=4$]
      { \includegraphics[width=\columnwidth]{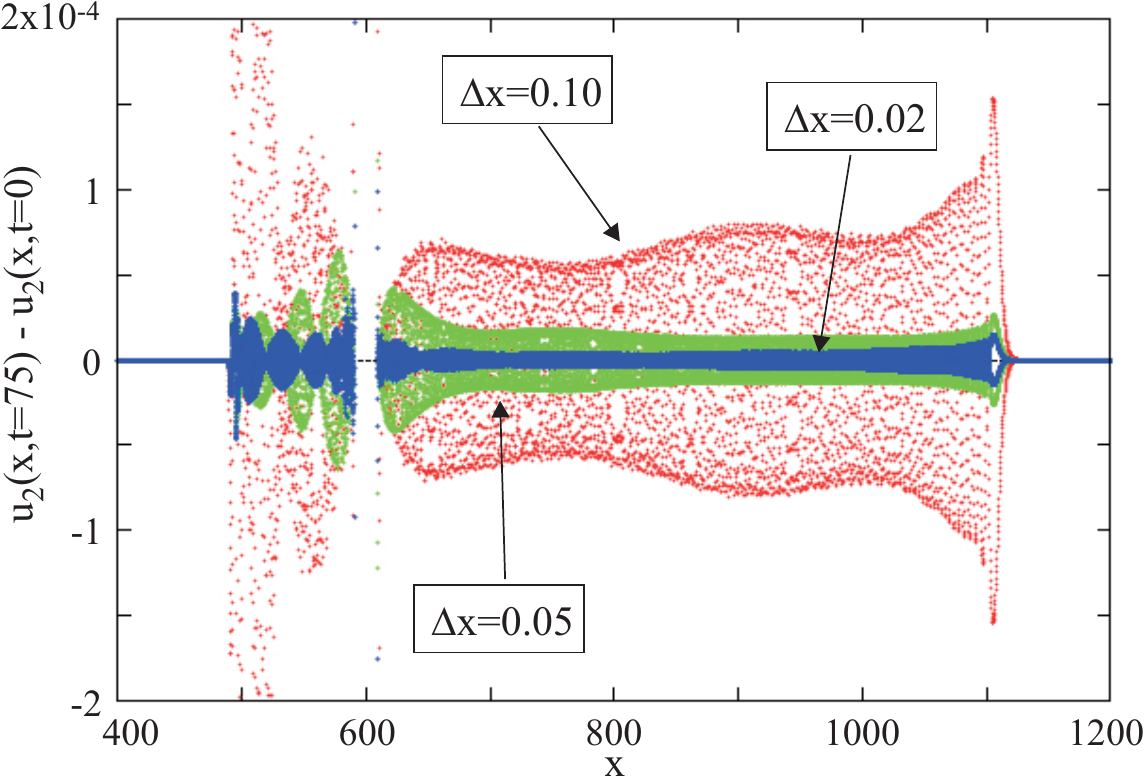} }
   \subfigure[CSS~$u_3$: $p=2,l=3$]
      { \includegraphics[width=\columnwidth]{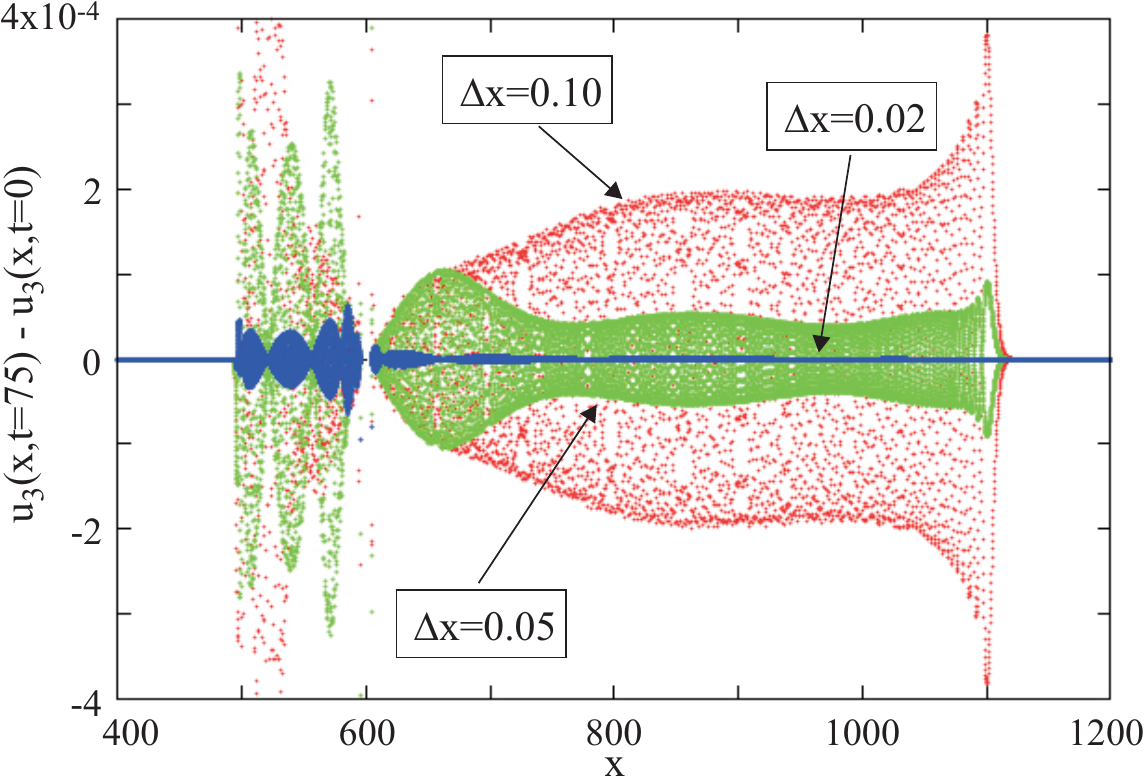} }
   \caption{\label{stab_dx} (Color online) Study of the CSS compacton stability. CSS results are compared with those obtained in the $K(2,2)$ case.
   Here, $u_1$ and $u_2$ are CSS compactons with velocity-independent width, corresponding to the case $l=p+2$, with $p=1$ and $p=2$, respectively, whereas
   $u_3$ is a CSS compacton with velocity-dependent width, corresponding to the case $p=2$, $l=3$.
   The numerically-induced radiation train results at time $t$=75 were obtained using the (6,4,4) scheme described in the text.
   The compactons were propagated in their comoving frames ($c_0=c$) with $\Delta t$=0.1 and $\Delta x$=0.1, 0.05, and 0.025. 
   In all cases, the radiation appears to be a numerical artifact that is suppressed by reducing the grid spacing, $\Delta x$. 
   This indicates that indeed these compactons are stable.}
\end{figure*}

%
%

\begin{figure*}[t]
   \subfigure[K(p,p) $u_c$: $p=2$.]
      { \includegraphics[width=\columnwidth]{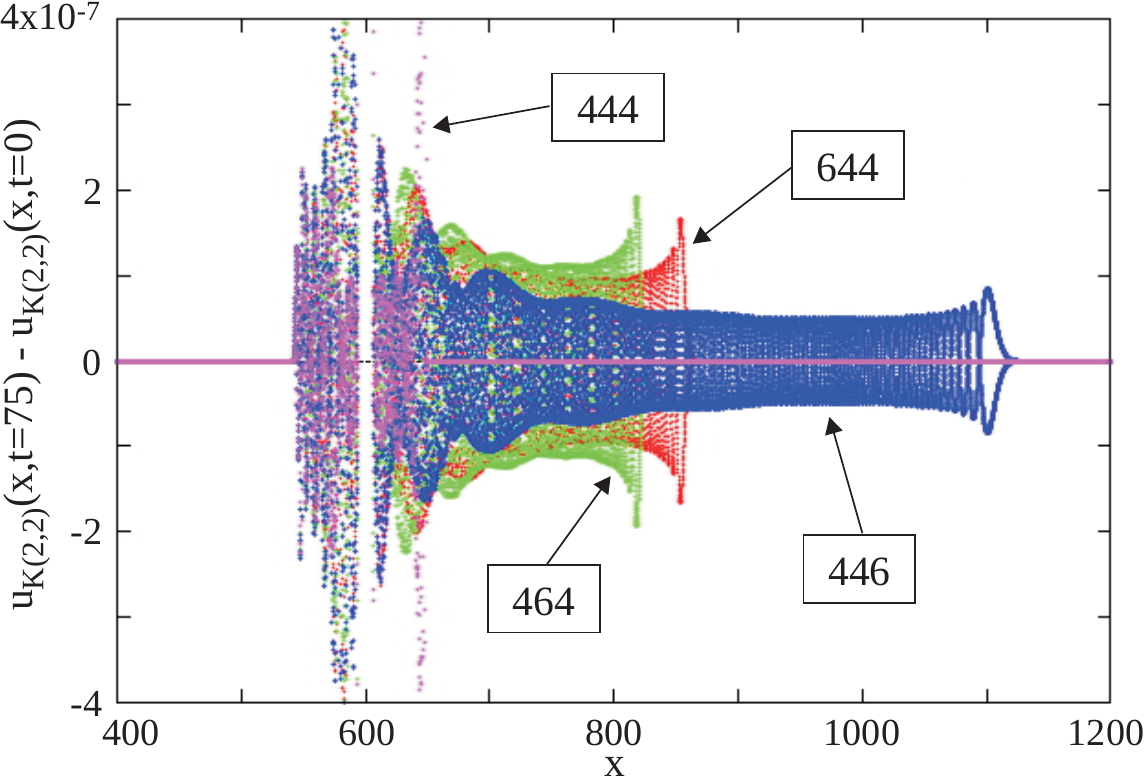} }
   \subfigure[CSS~$u_1$: $p=1,l=3$]
      { \includegraphics[width=\columnwidth]{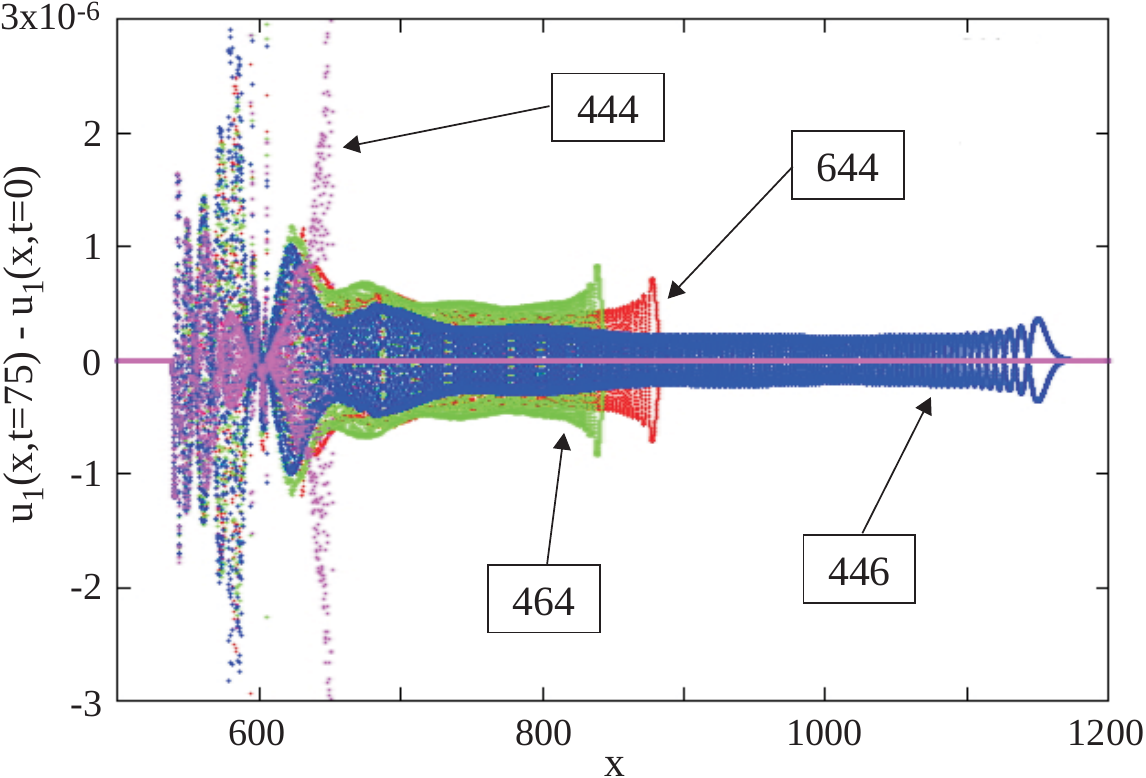} }
   \subfigure[CSS~$u_2$: $p=2,l=4$]
      { \includegraphics[width=\columnwidth]{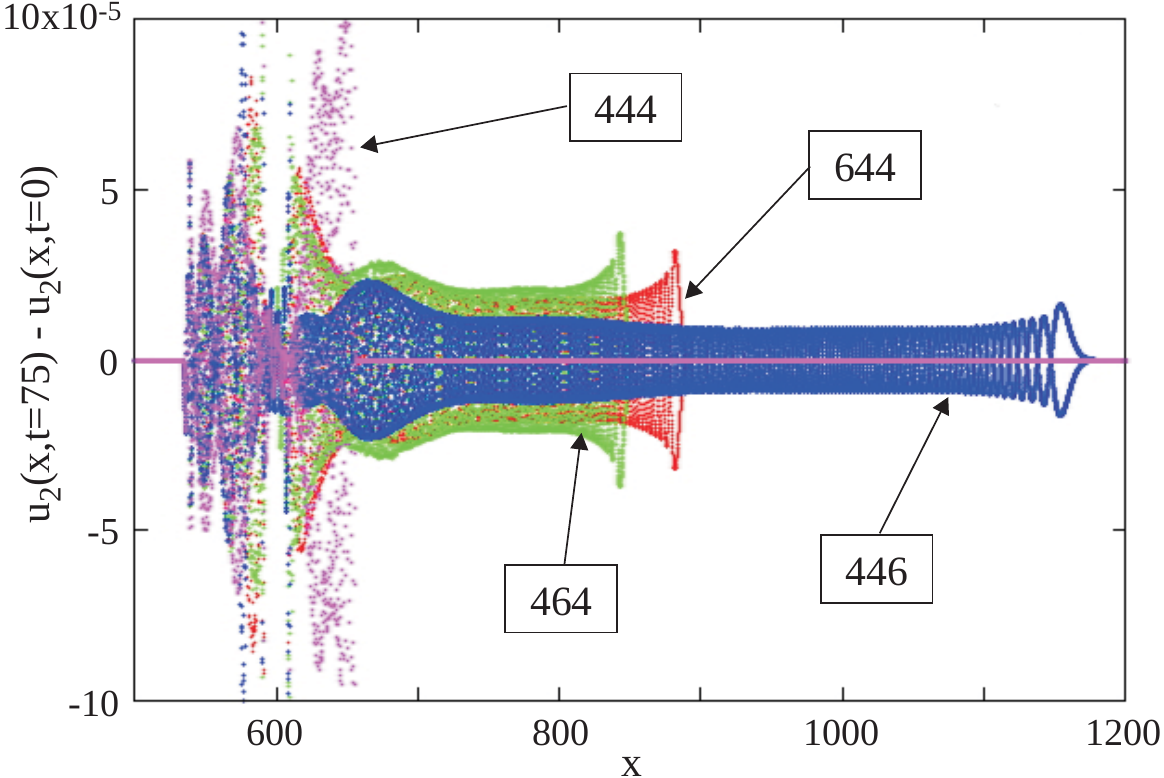} }
   \subfigure[CSS~$u_3$: $p=2,l=3$]
      { \includegraphics[width=\columnwidth]{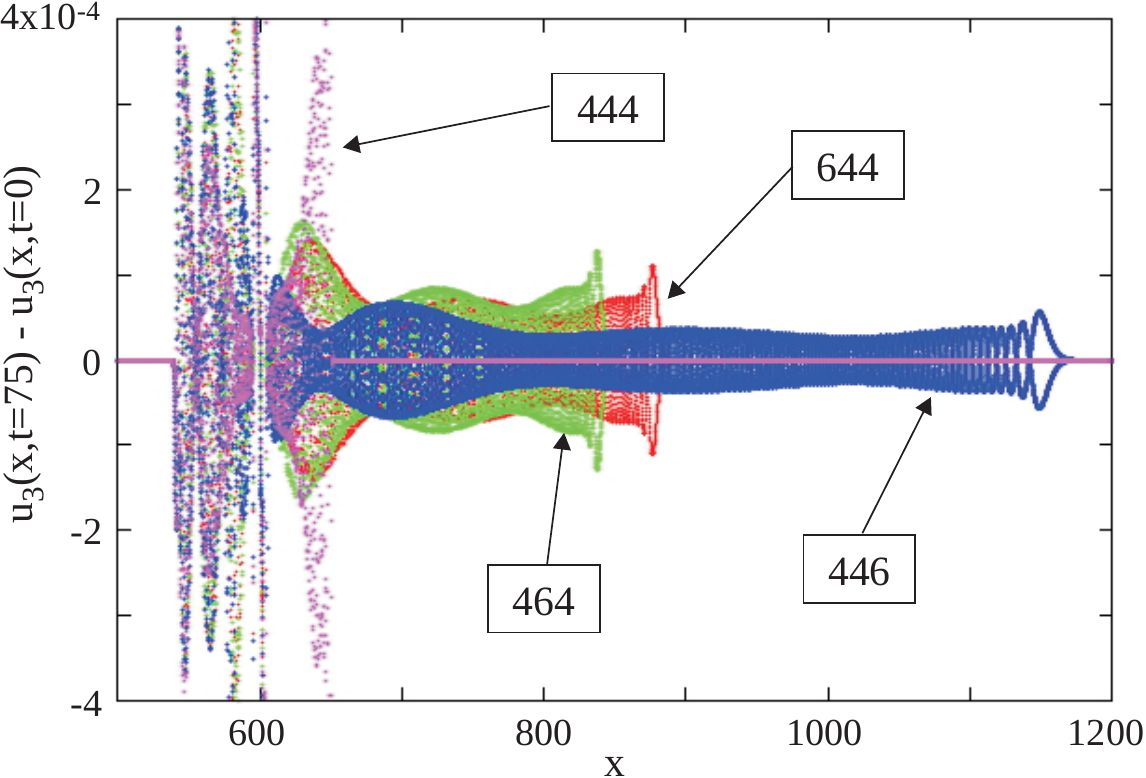} }
   \caption{\label{stab_methods} (Color online) Comparison of compacton stability results as a function of numerical scheme. 
   The CSS-compactons stability study is shown to be robust with respect to the chosen fourth-order accurate Pad\'e-approximant numerical scheme.
   CSS results are compared with those obtained in the $K(2,2)$ case.
   Here we plot the radiation trains obtained at $t$=75, by propagating the compactons in their comoving frames ($c_0=c$) with $\Delta t$=0.1 and $\Delta x$=0.05. 
   }
\end{figure*}

%
%

\section{Numerical approach}
\label{sec:numerics}


We will apply the numerical schemes based on the Pad\'e approximants discussed above to the case of the equation
\begin{align}
   \label{css_pt}
   u_t &
   - c_0 \, u_x
   + \eta \, u_{xxxx}
   + \frac{1}{l-1} \, \bigl ( u^{l-1} \bigr )_x
   \\ \notag & 
   - \alpha \, p \, \bigl ( u^{p-1} u_x^q \bigr )_x
   + \alpha \, q \, \bigl ( u^p u_x^{q-1} \bigr )_{xx}
   = 0
   \>,
\end{align}
where the subscripts $t$ and $x$ indicate partial derivatives with respect to $t$ and $x$, respectively.
Here, $u(x,t)$ is time evolved in the moving frame of reference with velocity $c_0$, and in the presence of an artificial dissipation (hyperviscosity) term based on fourth spatial derivative, $\eta \, \partial^4 u / \partial x^4$. The hyperviscosity term is needed to damp out 
explicitly the numerical high-frequency dispersive errors introduced by the lack of smoothness at the edge of the discrete representation of the compacton (see e.g. discussion in Ref.~\cite{CHK01}). The addition of artificial dissipation is responsible for the appearance of tails and compacton amplitude loss. In our dynamics simulations we choose $\eta$ as small as possible to reduce these numerical artifacts without significantly changing the solution to the compacton problem. We note that in the CSS-compacton simulations discussed here, we required a hyperviscosity value an order of magnitude smaller than the hyperviscosity used in our previous simulations of $K(2,2)$ compactons. Unless otherwise specified, we use $\eta = 10^{-5}$.

Setting $q=2$ in Eq.~\eqref{css_pt}, leads to the case of the CSS compacton derived from the Lagrangian~\eqref{eq:Llp}, i.e.
\begin{align}
   \label{css_num}
   u_t &
   - c_0 \, u_x
   + \eta \, u_{xxxx}
   + \frac{1}{l-1} \, \bigl ( u^{l-1} \bigr )_x
   \\ \notag & 
   - \alpha \, p \, \bigl ( u^{p-1} u_x^2 \bigr )_x
   +  \frac{2 \, \alpha}{p+1} \, \bigl ( u^{p+1} \bigr )_{xxx}
   = 0
   \>,
\end{align}
whereas for $\alpha=(q-1)^{-1}$ and $q$ an even integer, we obtain the compacton equation for the $\mathcal{PT}$-symmetric case discussed in Ref.~\onlinecite{BCKMS09}. Hence, even though in the following we focus on the discussion of the properties of the CSS compactons, the numerical methods developed here apply also to the case of $\mathcal{PT}$-symmetric compactons.

To obtain the spatial numerical discretization of Eqs.~\eqref{css_pt} and~\eqref{css_num}, suitable for our fourth-order accurate Pad\'e-approximant approach, we introduce a uniform spatial grid in the interval $x\in[0,L]$ by defining the grid points $x_m = m \Delta x$, with $m = 0,1,\cdots,M$ and the grid spacing $\Delta x = L/M$. Then, we have
\begin{align}
   0 = \ & 
   \mathcal{F}(E) \, \frac{\mathrm{d}u_m}{\mathrm{d}t}
   - 
   \bigl [ c_0 \mathcal{A}(E) - \eta \, \mathcal {D}(E) \bigr ] u_m
   \notag \\ & 
   +  \mathcal{A}(E)  \, 
   \Bigl [ \frac{1}{l-1} \, (u_m)^{l-1}
   - \alpha \, p \, (u_m)^{p-1} \, \bigl ( \{u_x\}_m \bigr )^q
   \Bigr ]
   \notag \\ & 
   + \alpha \, q \, \mathcal{B}(E)  \,  
   \Bigl [ (u_m)^{p} \, \bigl ( \{u_x\}_m \bigr )^{q-1} \Bigr ]
   \>.
\label{eq:1}
\end{align}
In Eq~\eqref{eq:1}, $u_m(t)$ is a numerical approximation to $u(x_m, t)$, and we assume that $u_m(t)$ obeys periodic boundary conditions, $u_{M}(t) = u_0(t)$. 
Also in Eq.~\eqref{eq:1}, we introduced the notation $u(x_m, t)$, and $\{u_x\}_m$ to denote a numerical approximation to $\partial_x u(x_m, t)$. The latter is calculated using Eq.~\eqref{til_A1}.
The optimal discretization for the study of CSS compactons corresponds to Eq.~\eqref{eq:1}. As such, for $q$=2, we have
\begin{align}
   0 = \ & 
   \mathcal{F}(E) \, \frac{\mathrm{d}u_m}{\mathrm{d}t}
   - 
   \bigl [ c_0 \mathcal{A}(E) - \eta \, \mathcal {D}(E) \bigr ] u_m
   \notag \\ & 
   +  \mathcal{A}(E)  \, 
   \Bigl [ \frac{1}{l-1} \, (u_m)^{l-1}
   - \alpha \, p \, (u_m)^{p-1} \, \bigl ( \{u_x\}_m \bigr )^2
   \Bigr ]
   \notag \\ & 
   + \frac{2 \, \alpha}{p+1} \, \mathcal{C}(E)  \,  
   \Bigl [ (u_m)^{p+1}\Bigr ]
   \>.
\label{eq:1css}
\end{align}

In order to numerically calculate the dynamics, we discretized the time-dependent parts of Eqs.~\eqref{eq:1} and~\eqref{eq:1css} in Eqs.~\eqref{css_pt} and~\eqref{css_num} by implementing the midpoint rule in time, similar to previous studies~\cite{pade_paper,RV09}. The resulting approximate equation for Eq.~\eqref{eq:1}~is
\begin{widetext}
\begin{align}
   0 = \ &
   \mathcal{F}(E) \, \frac{u_m^{n+1}-u_m^n}{\Delta t}
   - \Bigl [ c_0 \mathcal{A}(E) - \eta \mathcal {D}(E) \Bigr ] 
   \Bigl ( \frac{u_m^{n+1} + u_m^n}{2} \Bigr )
   + \frac{1}{(l-1)} \, \mathcal{A}(E)  \, 
   \Bigl ( \frac{u_m^{n+1} + u_m^n}{2} \Bigr )^{l-1}
\label{eq:1mid}
   \\ \notag & 
   - \alpha \, p \, \mathcal{A}(E)  \, 
   \Bigl [ 
   \Bigl ( \frac{u_m^{n+1} + u_m^n}{2} \Bigr )^{p-1} 
   \Bigl ( \frac{\{u_x\}_m^{n+1} + \{u_x\}_m^n}{2} \Bigr )^q 
   \Bigr ]
   + \alpha \, q \, \mathcal{B}(E)  \,  
   \Bigl [ 
   \Bigl ( \frac{u_m^{n+1} + u_m^n}{2} \Bigr )^{p} 
   \Bigl ( \frac{\{u_x\}_m^{n+1} + \{u_x\}_m^n}{2} \Bigr )^{q-1} 
   \Bigr ]
   \>.
\end{align}
Here we introduced the notations, $u_m^n=u_m(t_n)$ and $u_m^{n+1} =u_m(t_n+\Delta t)$, to indicate evaluations at two different moments of time.

For CSS compactons the discretization is
\begin{align}
   0 = \ &
   \mathcal{F}(E) \, \frac{u_m^{n+1}-u_m^n}{\Delta t}
   - \Bigl [ c_0 \mathcal{A}(E) - \eta \mathcal {D}(E) \Bigr ] 
   \Bigl ( \frac{u_m^{n+1} + u_m^n}{2} \Bigr )
   + \frac{1}{(l-1)} \, \mathcal{A}(E)  \, 
   \Bigl ( \frac{u_m^{n+1} + u_m^n}{2} \Bigr )^{l-1}
   \notag \\ & 
   - \alpha \, p \, \mathcal{A}(E)  \, 
   \Bigl [ 
   \Bigl ( \frac{u_m^{n+1} + u_m^n}{2} \Bigr )^{p-1} 
   \Bigl ( \frac{\{u_x\}_m^{n+1} + \{u_x\}_m^n}{2} \Bigr )^2
   \Bigr ]
   + \frac{2 \, \alpha}{p+1} \, \mathcal{C}(E)  \,  
   \Bigl ( \frac{u_m^{n+1} + u_m^n}{2} \Bigr )^{p+1}
   \>.
\label{eq:1mid_css}
\end{align}
\end{widetext}

%
%

\section{Results and discussion}
\label{sec:res}

\textcolor{black}{In the following, we discuss the case of the CSS compacton equation, given in the laboratory frame by Eq.~\eqref{eq:css} or, in a frame moving with velocity $c_0$, by Eq.~\eqref{css_num}, where in the latter we set the hyperviscosity to zero, $\eta$=0.} We study the properties of the three exact compacton solutions described in Ref.~\onlinecite{CSS93}. 
The first two of these compactons  correspond to a class of solutions with $l = p + 2$. The width of these compactons is independent of the compacton velocity, $c$, and the compactons have the general form
\begin{equation}
   u(x,t) = \Bigl [ \frac{c(p+1)(p+2)}{2} \Bigr ]^{\frac{1}{p}} 
   \cos^{\frac{2}{p}} \Bigl [ \frac{p \, \xi(x,t)}{\sqrt{4\alpha(p+1)(p+2)}} \Bigr ]
   \>,
\label{eq:lp2} 
\end{equation}
where we introduced the notation 
\begin{equation}
    \xi(x,t) = x - x_0 - (c-c_0)t
    \>,
\end{equation}
with $x_0$ the position of the compacton maximum at $t = 0$.
For $p = 1$ and $\alpha = \frac{1}{2}$, Eq.~\eqref{eq:lp2} gives the compacton solution
\begin{equation}
    u_1(x,t) = 3 \, c \ \cos^2 \Bigl [ \frac{1}{2 \sqrt{3}} \, \xi(x,t) \Bigr ]
    \>,
    \quad
    |\xi(x,t)| \le \sqrt{3}\pi
    \>,
\label{p1l3}
\end{equation}
whereas for $p=2$ and $\alpha=3$ we obtain the compacton solution
\begin{equation}
   u_2(x,t) = \sqrt{6 \, c} \ \cos \Bigl [ \frac{1}{6} \, \xi(x,t) \Bigr ]
   \>,
    \quad
    |\xi(x,t)| \le 3\pi
    \>.
\label{p2l4}
\end{equation}
The third compacton to be discussed next corresponds to the values, $p = 2$ and $l = 3$, and the width of this compacton depends on velocity. Choosing $\alpha = \frac{1}{4}$, we find
\begin{equation}
    u_3(x,t) = 3 \, c - \frac{1}{6} \ \xi^2(x,t)
    \>,
    \quad
    |\xi(x,t)| \le 3 \sqrt{2 \, c}
    \>.
\label{p2l3}
\end{equation}

\textcolor{black}{
Using Eq.~\eqref{eq:css_c}, one can study the possibility that the above compactons, $u_1$, $u_2$ and $u_3$, have anti-compacton counterparts. We infer that the CSS equation corresponding to the $u_1$ compactons ($p = 1$, $l = 3$) allows for anti-compacton counterparts traveling with a negative velocity, similar to the RH compactons. The $u_2$ equation ($p = 2$, $l = 4$) allows for compact solutions with negative amplitude, but these anti-compactons' have a positive velocity and travel in the same direction as $u_2$. Finally, the $u_3$ CSS equation ($p = 2$, $l = 3$) does not allow for anti-compacton solutions.
}

We will compare results of simulations for the above compacton solutions of the Lagrangian~\eqref{eq:Llp}, with results of similar simulations for compacton solutions of the RH generalization of the KdV equation, Eq.~\eqref{eq:kmn}: For $p$ restricted to the interval $1 < p \leq 3$, the $K(p,p)$ equation allows for compacton solutions of the form~\cite{IT98,RV07a,R98}
\begin{equation}
   u_c(x,t) = A^\gamma \cos^{2\gamma} \Bigl [ \beta \, \xi(x,t) \Bigr ]
   \>,
   \quad
   |\xi(x,t)| \leq \frac{1}{2 \, \beta} \, \pi \>,
\end{equation}
where
\begin{equation}
   A = \frac{2c \, p}{p+1} \>,
   ~\beta = \frac{p-1}{2p} \>,
   ~\gamma = \frac{1}{p-1} \>.
\end{equation}
For illustrative purposes, we will consider the case of the $K(2,2)$ equation ($p=2$), with the exact compacton solution
\begin{equation}
   u_c(x,t) = \frac{4c}{3} \cos^2 \Bigl [ \frac{1}{4} \, \xi(x,t) \Bigr ]
   \>,
   \quad
   |\xi(x,t)| \leq 2\pi
   \>.
\label{eq:uc}
\end{equation}

%
%

\subsection{Study of compacton stability}

To numerically demonstrate the stability of the CSS compacton solutions, we performed a numerical study of the compacton propagation in the compacton comoving frame ($c_0=c$), using the Pad\'e approximations discussed above, and we compare with results of similar simulations performed in the case of the $K(2,2)$ compacton that are known to be stable.

As shown in Fig.~\ref{stab_dx}, the numerical compactons propagate with the emission of forward and backward propagating radiation. If the compactons are numerically stable, then the amplitude of this radiation train is suppressed by reducing the grid spacing, $\Delta x$, which shows that the radiation train is a numerical artifact.
In Fig.~\ref{stab_dx}, we illustrate results obtained with the (6,4,4) Pad\'e-approximant scheme. Here we chose a snapshot at  $t$=75 after propagating the compacton in the absence of hyperviscosity ($\eta$=0) with a time step, $\Delta t$=0.1, and grid spacings, $\Delta x$=0.1, 0.05, and 0.025. The amplitude of the radiation train is at least 4~orders of magnitude smaller than the amplitude of the compacton. Using the grid refining technique, we can show that indeed the radiation is a numerically-induced phenomenon. The noise is suppressed by reducing the grid spacing, $\Delta x$, indicating that all studied CSS compacton~\eqref{eq:uc} solutions are stable.

These results are robust with respect to the choice of the fourth-order accurate Pad\'e-approximant numerical scheme. As shown in Fig.~\ref{stab_methods}, the extent and amplitude of the radiation train is a characteristic of the chosen numerical scheme, and results for the CSS compactons are ``identical" with results obtained for the K(2,2) compactons, albeit for a scaling in the amplitude of the radiation for a given choice of the time step ($\Delta t$=0.1) and grid spacing ($\Delta x$=0.05). This scaling is indicative of the higher nonlinearity of the CSS equation relative to the $K(2,2)$ equation\textcolor{black}{, as observed also when one compares the results for the $K(2,2)$ and $K(3,3)$ equations~\cite{RV07b}.
}

\textcolor{black}{We note that the origin of the radiation observed in the propagation of a compacton was shown previously to be of numerical origin in the case of the $K(p,p)$ equation by Rus and Villatoro~\cite{RV07b}, who also showed that this self-similarity depends strongly on the time-integration method~\cite{RV10}. The induced radiation depicted in Figs.~\ref{stab_dx} and~\ref{stab_methods} is similar to that of Ref.~\cite{RV07b} and therefore one would expect that the self-similarity of the radiation is also a feature of the CSS equation.}

%
%

\begin{figure}[t]
   \centering
   \includegraphics[width=\columnwidth]{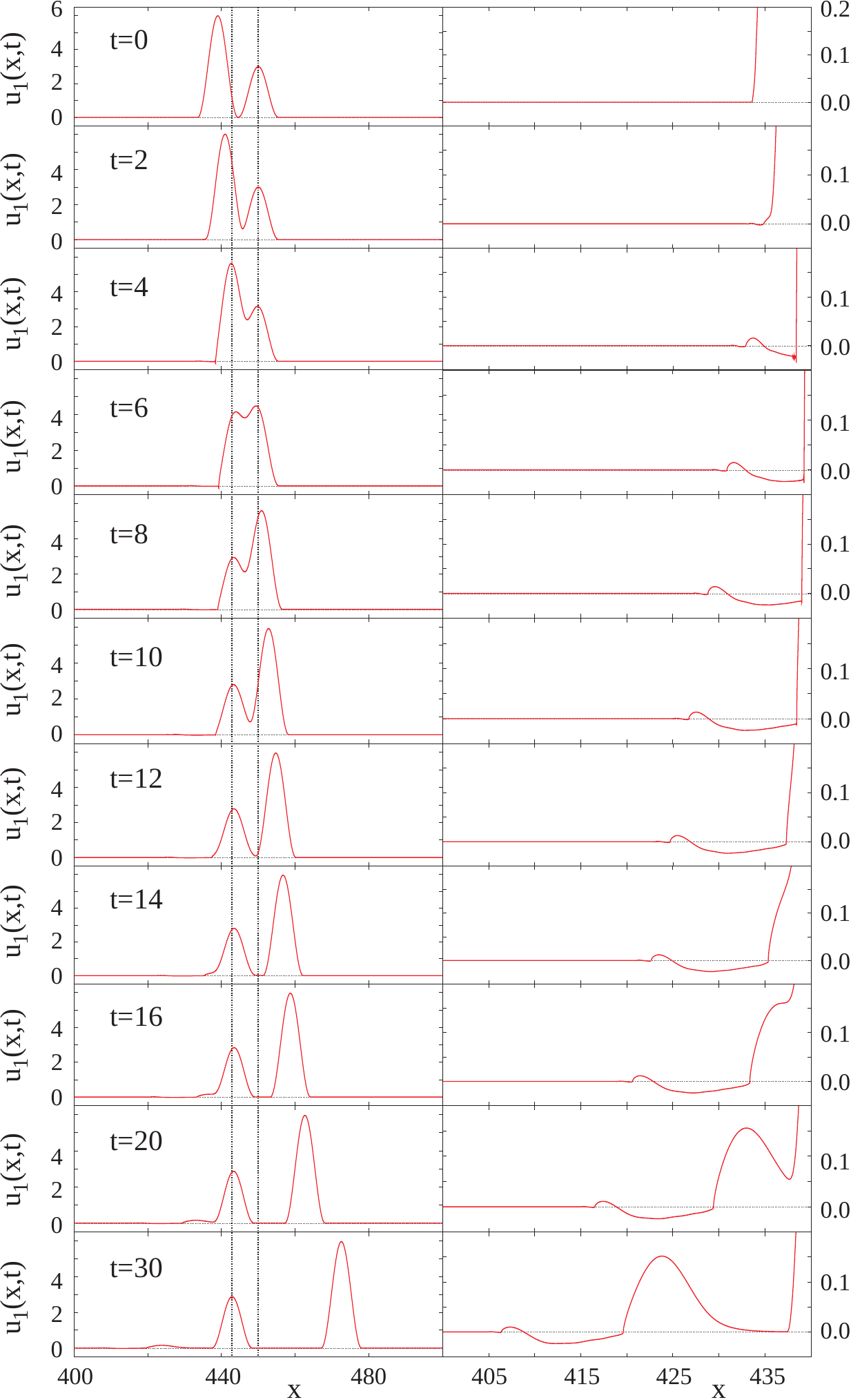}
   \caption{\label{p1l3_movie}(Color online)
   Collision of two CSS compactons, $u_1(x,t)$, with $c_1=1$ and $c_2=2$. 
   The width of $u_1$ compactons is independent of the compacton velocity and they correspond to the choice of parameters, $p=1$ and $l=3$.
   The simulation is performed in the comoving frame of reference of the first compacton, i.e. $c_0=c_1$, using the (6,4,4) scheme and a 
   hyperviscosity, $\eta=2\times10^{-5}$. 
   In the left panels, the collision is shown to be inelastic, despite the fact that the compactons maintain their coherent shapes after the collision:
    The first compacton ($c_1$=1) is ``at rest'' before the collision occurs. 
    As shown in the left panels, after the collision the centroid of this compacton changes position.
   The right panels depict the early development of the ripple created in the collision process.}
\end{figure}

\begin{figure}[t]
   \centering
   \includegraphics[width=0.8\columnwidth]{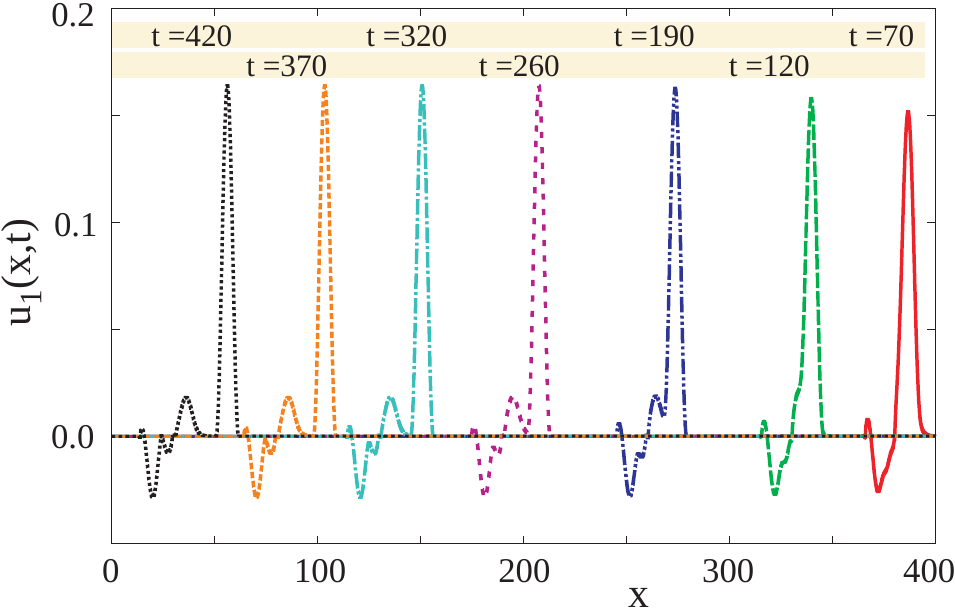}
   \caption{\label{p1l3_shake}(Color online)
   At intermediate times, the ripple created as a result of the collision of the two $u_1(x,t)$ CSS compactons depicted in Fig.~\ref{p1l3_movie} leads to the emergence of the first compacton at $t\sim400$. The decay process is very sluggish, similar to the RH~case.}
\end{figure}

\begin{figure}[t!]
   \centering
   \includegraphics[width=0.9\columnwidth]{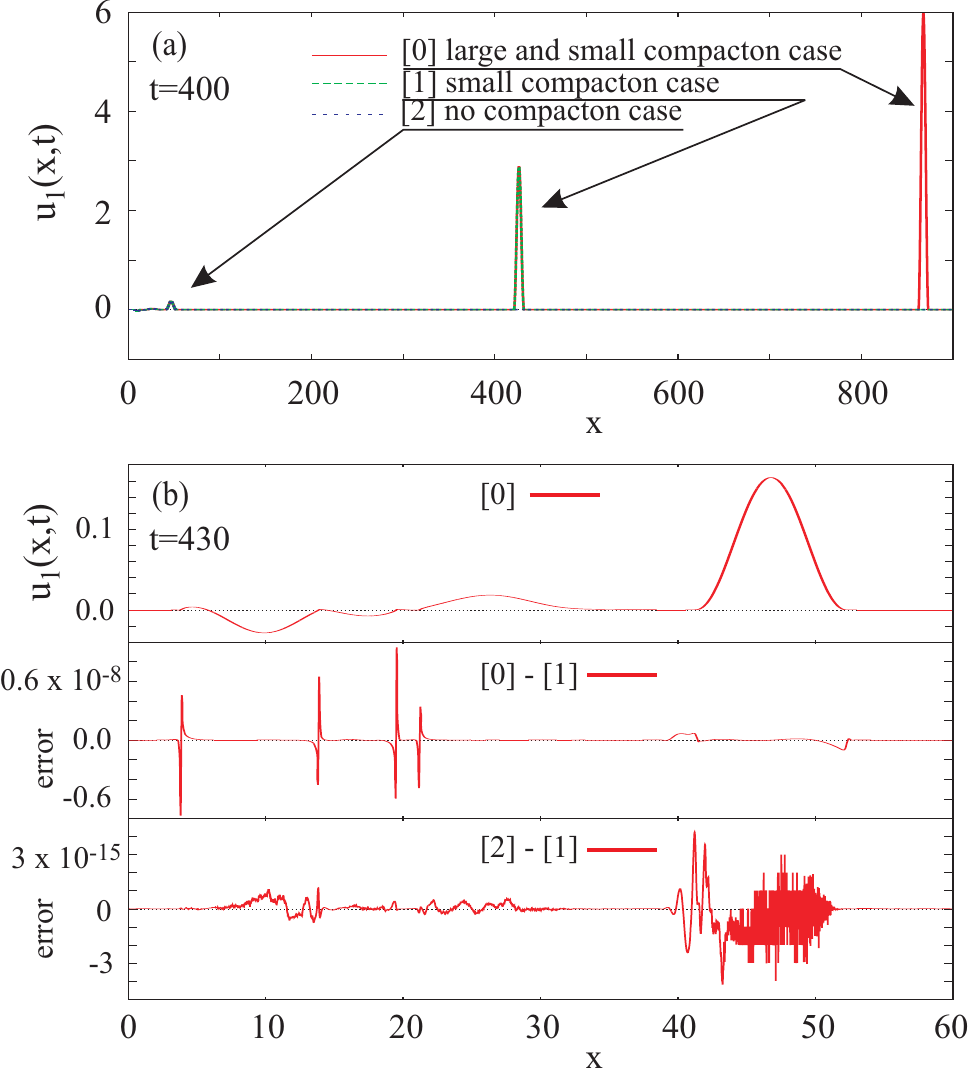}
   \caption{\label{p1l3_diff}(Color online)
   We study the correlations between the ripple and the two reemerging compactons: we stop the simulation of the collision process at time $t=400$, see plot labelled [0] in panel~(a). We use this snapshot to initialize two additional simulations: [1] a simulations in which we drop the large compacton, and [2] a simulation in which we drop both compactons. In panel~(b), we compare results of the three simulations  at $t=430$. We show that the differences between [0] and [1] are lower than the order of magnitude of the noise induced by the numerical discretization of the problem (compare with the noise depicted in Fig.~\ref{stab_dx}. The differences between [1] and [2] are of the order of the machine precision errors. The above indicate a lack of correlations between the ripple and the reemerging compactons.}
\end{figure}

\begin{figure}[t]
   \centering
   \includegraphics[width=0.9\columnwidth]{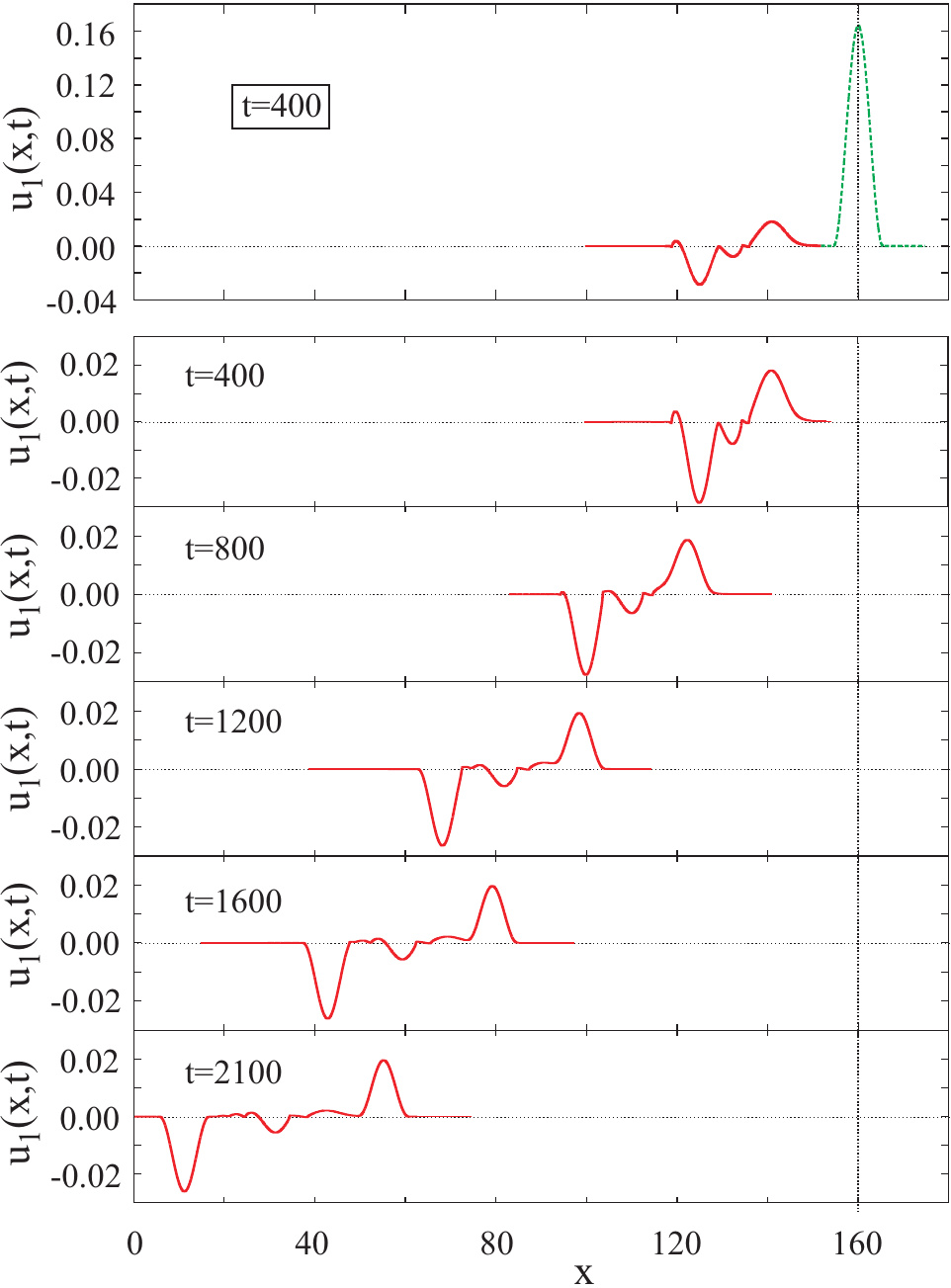}
   \caption{\label{p1l3_long}(Color online)
   Study of the ripple decomposition dynamics at late times. 
   Here we assumed that the correlations between the ripple and the two reemerging compactons are negligible, as indicated in Fig.~\ref{p1l3_diff}.
   For illustrative purposes, we shift the ripple, such that the position of the first emerging compacton from the ripple, at $t=400$, is kept fixed (upper panel). In the snapshots depicted in the lower panels, we observe the emergence of anti-compacton in the $t=1200$ graph, and the emergence of a second compacton at $t=2100$. }
\end{figure}

\begin{figure}[t]
   \centering
   \includegraphics[width=\columnwidth]{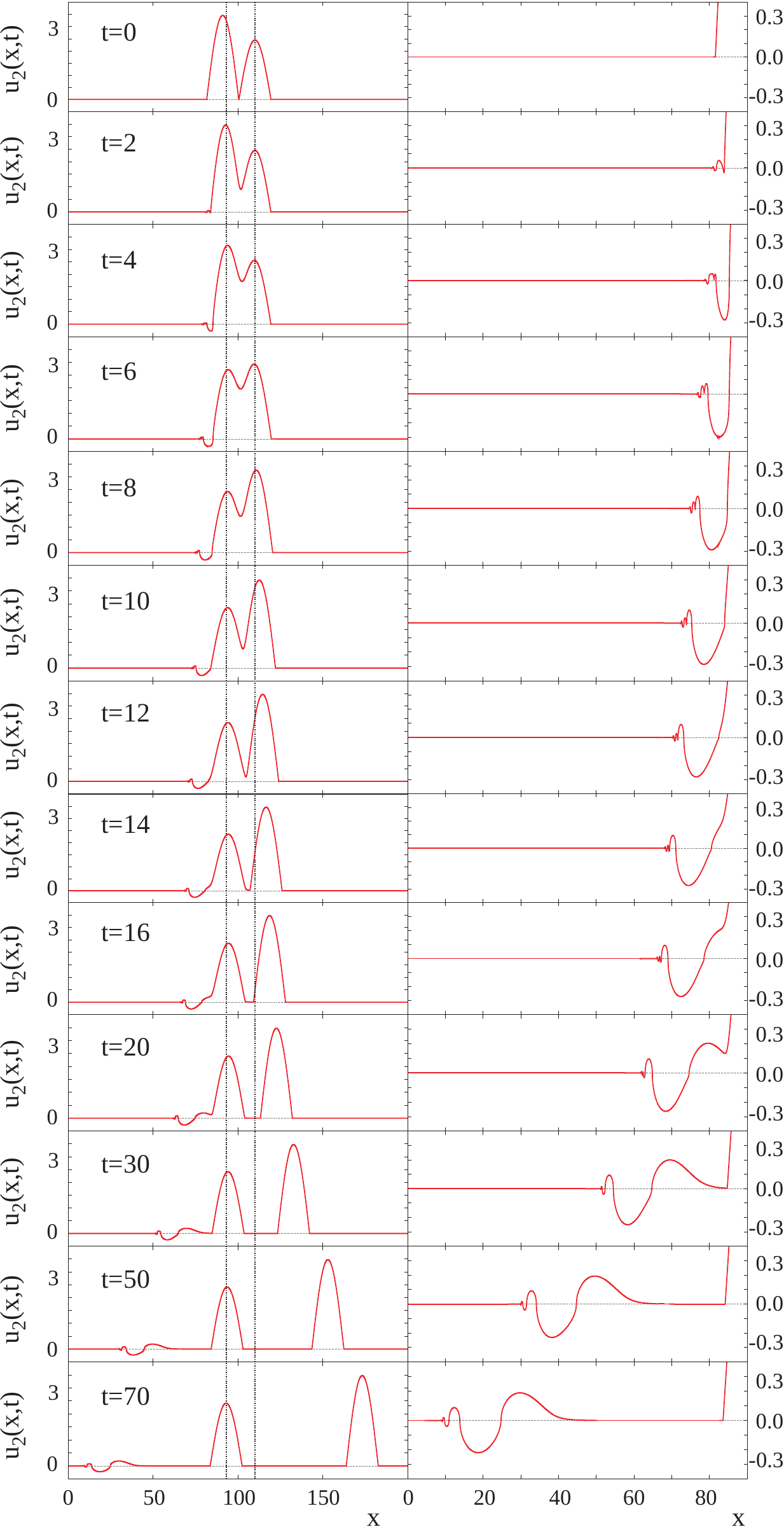}
   \caption{\label{p2l4_movie}(Color online)
   Collision of two $u_2(x,t)$ compactons, with $c_1=1$ and $c_2=2$. 
   The width of $u_2$ compactons is also independent of the compacton velocity and they correspond to the choice of parameters, $p=2$ and $l=4$.
   The simulation is performed in the comoving frame of reference of the first compacton, i.e. $c_0=c_1$, using the (6,4,4) scheme and a 
   hyperviscosity, $\eta=10^{-5}$. Results are similar to the ones depicted in Fig.~\ref{p1l3_movie}.
   }
\end{figure}

\begin{figure}[t]
   \centering
   \includegraphics[width=\columnwidth]{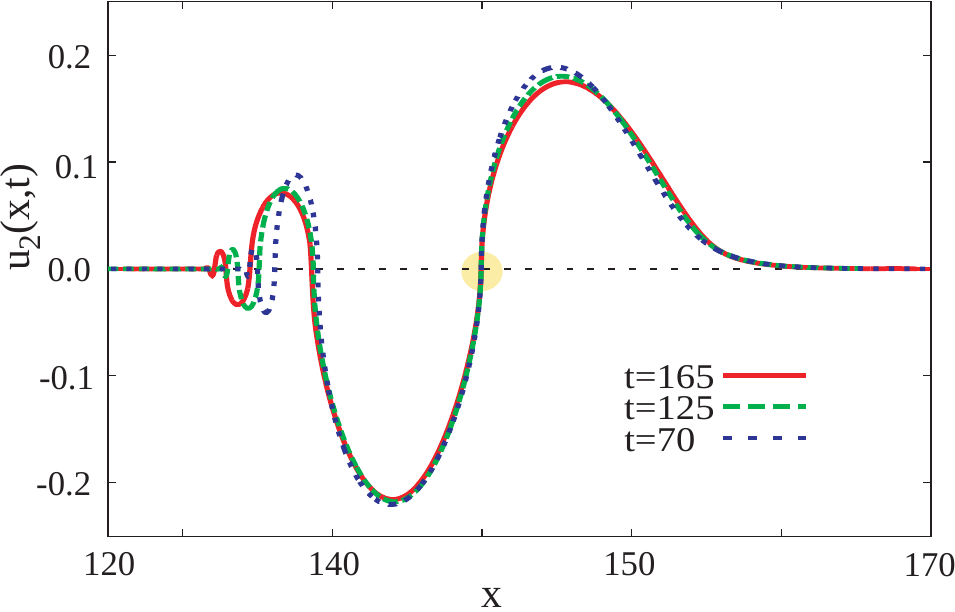}
   \caption{\label{p2l4_comp}(Color online)
   Dynamics of the ripple resulting from the collision of two $u_2(x,t)$ compactons depicted in Fig.~\ref{p2l4_movie}. 
   We compare the shape of the ripple at different times in the time propagation, after the ripple ``separated" from the compactons.
   In order to compare the shapes of the ripple at different times, we forced the ripples to cross the $x$~axis 
   at the point indicated in the figure.
   Therefore the $x$~coordinates indicated here are only intended to indicate the spatial extent of the ripple.
   }
\end{figure}

\begin{figure}[b]
   \centering
   \includegraphics[width=0.8\columnwidth]{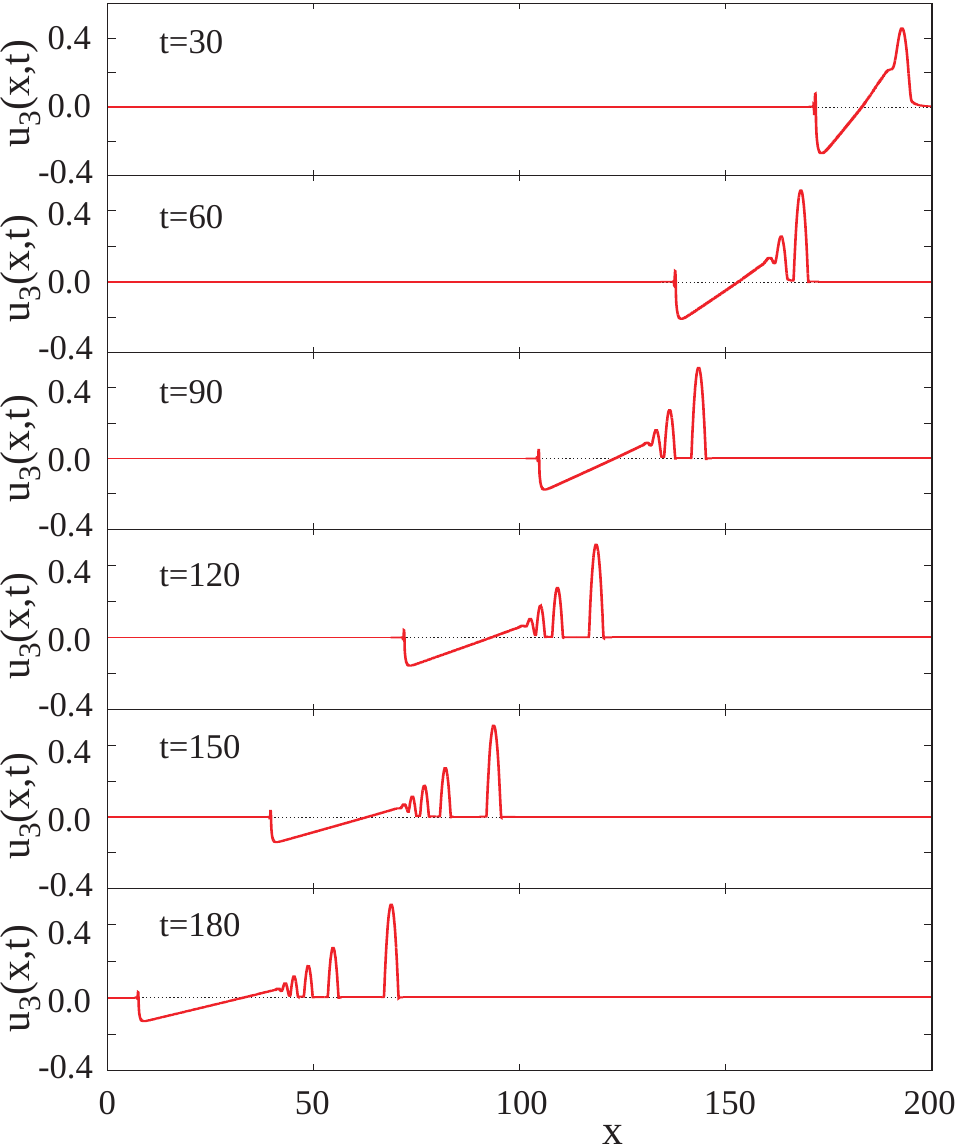}
   \caption{\label{p2l3_shake}(Color online)
   Dynamics of the ripple created as a result of the collision of two CSS $u_3(x,t)$ compactons. 
   The width of $u_3$ compactons depends on the compacton velocity and they correspond to the choice of parameters, $p=2$ and $l=3$.
   The simulation is performed in the comoving frame of reference of the first compacton, i.e. $c_0=c_1$, using the (6,4,4) scheme and a 
   hyperviscosity, $\eta=10^{-5}$.
   We note that the ripple ``decays'' in a suite of compactons, without any anti-compacton counterparts. 
   The dynamics of this process is much faster than in the case of collision between RH compactons or CSS compactons 
   with compacton velocity-independent widths.}
\end{figure}

%
%

\subsection{Pairwise interaction of CSS compactons}

In the following we will show that the CSS compactons also have the soliton property of remaining intact after the collisions. \textcolor{black}{The ripple generated following the reemergence of the CSS compactons decomposes into compactons with or without anti-compacton counterparts, depending on the values of the $l$ and $p$ parameters in the corresponding CSS equation.
In this context, it is important to recall that the $u_1$ compactons are the only CSS compactons that have anti-compacton counterparts traveling with a negative velocity, similar to the RH compactons. The $u_2$ CSS equation allows for anti-compacton solutions with negative amplitude, but with positive velocity, traveling in the same direction as $u_2$. Finally, the $u_3$ CSS equation does not allow for anti-compacton solutions.
}
No evidence of shock formation accompanying the collision was observed.

All simulations described next involve collisions between two CSS compactons with velocities $c_1=1$ and $c_2=2$. The compactons are  propagated in the comoving frame of reference of the first compacton, i.e. $c_0=c_1$, using the (6,4,4) Pad\'e approximant scheme. All simulations were performed in the presence of an artificial hyperviscosity. Unless otherwise stated, the hyperviscosity value was $\eta=10^{-5}$, an order of magnitude less than the hyperviscosity used in our previous simulations of the $K(2,2)$ compacton collisions~\cite{pade_paper}. 

We consider first the collision between two CSS $u_1(x,t)$ compactons,  see Eq.~\eqref{p1l3}, with parameters $p=1$ and $l=3$. The width of the $u_1$ compactons is independent of the compacton velocity. 

\textbf{Case~1: $p=1$,~$l=3$.} In Fig.~\ref{p1l3_movie}, we depict a series of snapshots of this collision process.  Just like in the $K(2,2)$ compacton case, the collision is shown to be inelastic, despite the fact that the compactons maintain their coherent shapes after the collision. The first compacton is ``at rest" before the collision occurs. As shown in the left panels of Fig.~\ref{p1l3_movie}, after collision this compacton  emerges with the centroid located at a new spatial position. The early development of the ripple created as a result of the pairwise compacton collision is illustrated in the right panels of Fig.~\ref{p1l3_movie}. 

In Fig.~\ref{p1l3_shake} we illustrate the emergence of the \emph{first} compacton from the ripple. We note the very sluggish decay process, just like in the RH-compacton case~\cite{RH93}.
   
To demonstrate the lack of correlations between the ripple and the two reemerged compactons after collision,  we use the result of the simulation at $t=400$ [denoted as [0] in panel~(a) of Fig.~\ref{p1l3_diff}], to initialize two additional simulations: [1] a simulation in which we drop the large compacton, and [2] a simulation in which we drop both compactons. In panel~(b) of Fig.~\ref{p1l3_diff}, we compare results of the three simulations  at $t=430$. Here, we note that the differences between [0] and [1] are lower than the order of magnitude of the noise induced by the numerical discretization of the problem (e.g. compare with the noise depicted in Fig.~\ref{stab_dx}). The differences between [1] and [2] are of the order of the machine precision errors.

\textcolor{black}{Assuming that the correlations between the ripple and the two reemerging compactons are negligible, we can study the dynamics of the ripple at late times. For illustrative purposes, we shift the ripple, such that the position of the first emerging compacton is kept fixed. In Fig.~\ref{p1l3_long}, we depict  snapshots of the ripple decomposition for $t \le 2100$. Here, we note the emergence of a first anti-compacton in the $t=1200$ graph, and the emergence of a second compacton at $t=2100$. In the $p=1$ and $l=3$ case, the emerging compactons and anti-compactons are moving in opposite directions relative to the remaining ripple, which is considerably reduced in amplitude.
}

\textbf{Case~2: $p=2$,~$l=4$.} Similar to the collision process depicted in Fig.~\ref{p1l3_movie}, in Fig.~\ref{p2l4_movie} we present a series of time snapshots illustrating the collision of two CSS $u_2(x,t)$ compactons. The width of the $u_2$ compactons is also independent of the compacton velocity and these compactons correspond to the choice of parameters $p=2$ and $l=4$. The results depicted in Fig.~\ref{p2l4_movie} are similar to those in Fig.~\ref{p1l3_movie}, albeit for the differences in the shape of the emerging ripple. 

The dynamics of the $u_2$  ripple is illustrated in Fig.~\ref{p2l4_comp}.   In order to compare the shapes of the ripple at different times after the ripple ``separated" from the reemerging compactons, in Fig.~\ref{p2l4_comp} we plot them such that they all cross the $x$~axis at the point indicated in the figure. The shape of the ripple is shown to be evolving very slowly, \textcolor{black}{likely as a result of the fact that in the $u_2$ case compacton and anti-compacton solutions travel in the same direction}. As going to later times in this simulation was deemed too expensive computationally, we chose to terminate it  \emph{before} any compacton \textcolor{black}{or anti-compacton} emerged from the ripple.

\textbf{Case~3: $p=2$,~$l=3$.} In Fig.~\ref{p2l3_shake}, we illustrate the dynamics of the ripple created as a result of  the collision of two CSS $u_3(x,t)$ compactons.  The $u_3$ compactons correspond to parameters, $p=2$ and $l=3$, and their widths depend on the compacton velocity. We note that the ripple ``decays'' in a suite of compactons, without any anti-compacton counterparts, \textcolor{black}{as the CSS equation for $p=2$ and $l=3$ does not allow for anti-compacton solutions. The amplitude of the ripple in this case is much larger than in the case of collisions between RH compactons or CSS compactons with compacton velocity-independent widths, and this may explain why the dynamics of the ripple-decomposition process is much faster in the $p=2$ and $l=3$ case.}

%
%

\section{Conclusions}
\label{sec:concl}

To summarize, in this paper we presented a systematic study of the stability and dynamical properties of CSS compactons. Several numerical schemes based on fourth-order Pad\'e approximants have been employed and the results were found to be independent of the numerical scheme. 
We find that  for the propagation of the CSS compactons in time 
using the implicit midpoint rule leads to stable results. 
The simulation of the CSS compacton scattering requires a much smaller artificial viscosity to obtain numerical stability, than in the case of RH compactons propagation. 

Based on our study, we verified numerically the conclusion of stability regarding the CSS compactons first derived based on criteria such as Lyapunov stability~\cite{Lyapunov}  and stability of the solutions under scale transformations~\cite{Derrick}. 

Just like in the case of  RH compactons, the CSS compactons preserve their  coherent shapes after the collision. 
\textcolor{black}{The ripple generated following the reemergence of the CSS compactons depends on the values of the parameters $l$ and $p$ characterizing the CSS compactons:
For a given set of $l$ and $p$ values, the ripple decomposition gives rise to compactons and anti-compacton counterparts, depending on the presence and character of the anti-compacton solutions allowed by the corresponding CSS equation.
}
The decomposition of the ripple is much faster for a class of CSS compactons for which the width of the compacton depends on its velocity. 
No evidence of shock formation accompanying the collision was observed after the collisions between CSS compactons.

\begin{acknowledgments} 
   This work was performed in part under the auspices of the United States Department of Energy.  
   B. Mihaila and F. Cooper would like to thank the Santa Fe Institute for its hospitality during the completion of this work.
\end{acknowledgments}

\vfill


%
%

\end{document}